\documentclass[aip, jcp, preprint,showpacs,superscriptaddress,groupedaddress]{revtex4-1}

\usepackage{media9}

\usepackage[vcentermath]{youngtab}
\usepackage{young}
\usepackage{amsmath}
\usepackage{amssymb}
\usepackage{braket}
\usepackage{bm}
\usepackage{tikz}
\usetikzlibrary{decorations.pathmorphing}

\usepackage{subcaption}
\usepackage{colortbl}

\begin{document}

\title{Richardson-Gaudin states of non-zero seniority II: Single-reference treatment of strong correlation}
\author{Paul A. Johnson}
 \email{paul.johnson@chm.ulaval.ca}
 \affiliation{D\'{e}partement de chimie, Universit\'{e} Laval, Qu\'{e}bec, Qu\'{e}bec, G1V 0A6, Canada}

\date{\today}

\begin{abstract}
Strongly correlated systems are well described as a configuration interaction of Slater determinants classified by their number of unpaired electrons. This treatment is however unfeasible. In this manuscript, it is demonstrated that single reference methods built from Richardson-Gaudin states yield comparable results at polynomial cost.
\end{abstract}

\maketitle

\section{Introduction}
Most systems in chemistry are weakly correlated. Their behaviour is described by \emph{one} Slater determinant and corrections may be added systematically to improve the result. In strongly correlated systems it is difficult to find an adequate description in this picture. While the complete active space self-consistent field (CASSCF)\cite{roos:1980,siegbahn:1980,siegbahn:1981,roos:1987} is standard, it is not easy to use, and does not always lead to a clear understanding. Full configuration interaction (CI) is always correct, but unfeasible, so selective CIs\cite{huron:1973,sharma:2017,holmes:2017,li:2018,yao:2021} aim to pick the important Slater determinants automatically. In general this is difficult. Larger systems have been treated with the density matrix renormalization group (DMRG).\cite{chan:2002,ghosh:2008,yanai:2009,wouters:2014,sun:2017,ma:2017}

For strongly correlated systems it is productive to classify Slater determinants by their number of unpaired electrons, their \emph{seniority}. It was clearly established in ref. \citenum{bytautas:2011} that a CI based on seniority converges quickly, even for the nitrogen molecule. Seniority-zero CI, or historically doubly-occupied CI (DOCI), is qualitatively correct and more or less parallel to full CI. Adding the Slater determinants of seniority-two improves the result substantially, while including all Slater determinants up to seniority-four gives a result that is near-exact. However, there are two difficulties that prevent seniority-based CI from being practical. First, seniority is not conserved: a Slater determinant that has zero seniority in one set of orbitals could have a different seniority in another set of orbitals. The orbitals must therefore be optimized. While this is difficult in general, often the optimal orbitals are localised and reminiscent of valence-bond orbitals.\cite{limacher:2014a} Effective guesses for these orbitals may be constructed systematically,\cite{wang:2019,aldossary:2022} and hence this is not the present focus. The larger problem is that even seniority-zero CI has a cost that grows faster than exponentially. It was quickly discovered that pair coupled-cluster doubles (pCCD) was able to reproduce seniority-zero CI with mean-field cost.\cite{limacher:2013,limacher:2014b,boguslawski:2014a,boguslawski:2014b,boguslawski:2014c,tecmer:2014,stein:2014,henderson:2014a,henderson:2014b} Building on top of pCCD, or other seniority-zero approximations, is not straightforward though attempts have been made.\cite{cassam:2006,parkhill:2009,parkhill:2010,cassam:2010,boguslawski:2015,boguslawski:2016a,boguslawski:2016b,lehtola:2016,boguslawski:2017,lehtola:2018,boguslawski:2019,vu:2019,boguslawski:2021,marie:2021,kossoski:2021,baran:2021,cassam:2023,johnson:2025a}

As an alternative to Slater determinants, Richardson\cite{richardson:1963,richardson:1964,richardson:1965}-Gaudin\cite{gaudin:1976,gaudin_book} (RG) states describe weakly correlated \emph{pairs} of electrons. Systems which are strongly correlated in terms of Slater determinants are weakly correlated in RG states: seniority-zero CI is equivalent to one RG state plus a single-reference Epstein\cite{epstein:1926}-Nesbet\cite{nesbet:1955} perturbation theory (ENPT) correction.\cite{johnson:2024b} The next step is to determine whether this remains true in higher seniorities. Matrix elements between RG states of seniorities zero, two and four were computed in exhaustive detail in ref. \citenum{johnson:2025b}, hereafter referred to as Part I. As the matrix elements are rather complicated, details in the present contribution will be kept to an absolute minimum. Tt will be demonstrated that a short CI expansion based on a single RG reference reproduces the seniority-based CI curves for linear H$_8$ in particular. 

\section{Richardson-Gaudin States}
RG states are defined as the eigenvectors of the reduced Bardeen-Cooper-Schrieffer (BCS)\cite{bardeen:1957a,bardeen:1957b,schrieffer_book} Hamiltonian
\begin{align}
	\hat{H}_{BCS} = \frac{1}{2} \sum^N_{i=1} \varepsilon_i \left( a^{\dagger}_{i\uparrow} a_{\i\uparrow} + a^{\dagger}_{i\downarrow} a_{i\downarrow}\right)
	- \frac{g}{2} \sum^N_{i,j=1} a^{\dagger}_{i\uparrow}a^{\dagger}_{i\downarrow} a_{j\downarrow} a_{j\uparrow}
\end{align}
in terms of second quantized operators $a^{\dagger}_{i\uparrow}$ ($a_{i\downarrow}$) which create (remove) an up-spin (down-spin) electron in the spatial orbital $i$. This Hamiltonian expresses competition between an aufbau filling of spatial orbitals with energies $\varepsilon_i$ and a constant-strength pair scattering between them. It is exactly solvable for any values of the parameters $\{\varepsilon,g\}$. The RG states for $2M$ electrons in $N$ orbitals may be built from the eigenvalue-based-variables (EBV) $\{V\}$ which are solutions of the non-linear equations
\begin{align} \label{eq:ebv}
	V^2_i - 2V_i - g \sum^N_{j (\neq i)=1} \frac{V_j - V_i}{\varepsilon_j - \varepsilon_i} = 0, \qquad \forall i=1,\dots,N.
\end{align}
These equations are easily solved\cite{faribault:2011,elaraby:2012} by evolving them from $g=0$, where they decouple to
\begin{align}
	V_i (V_i - 2) = 0.
\end{align}
At $g=0$, there is no interaction, and hence the RG states reduce to closed-shell Slater determinants defined by sets of orbitals which are either doubly-occupied ($V_i=2$) or empty ($V_i=0$). These may be summarized as a bitstring with a 1 (0) indicating a doubly-occupied (empty) orbital. Further, these states evolve continuously from $g=0$, so that at any finite value of $g$, the bitstring labels continue to be meaningful. Seniority-two RG states are similar, with the added complication that two particular orbitals are each singly-occupied. These orbitals are \emph{blocked}. The corresponding EBV for the state are solutions of $N-2$ equations \eqref{eq:ebv} with contributions from the blocked orbitals removed. These states are labelled as bitstrings, augmented with Xs denoting which orbitals are blocked. Similarly, seniority-four RG states have four blocked orbitals, and the EBV are solutions of $N-4$ equations \eqref{eq:ebv} which have no contribution from their blocked orbitals. The final complication is that while there are three seniority-four singlets, only two are linearly independent, so that a choice must be made. The EBV for both seniority-four singlets are identical.

Matrix elements involving RG states are computable directly from their EBV. In particular, for two RG states with EBV $\{U\}$ and $\{V\}$, all that is required are the cofactors of the matrix
\begin{align}
	J_{ij} = \begin{cases}
		U_i + V_i - 2 \sum^N_{k (\neq i)=1} \frac{g}{\varepsilon_k - \varepsilon_i}, & i=j, \\
		\frac{g}{\varepsilon_i - \varepsilon_j}, & i \neq j.
	\end{cases}
\end{align}
For seniority-zero RG states, this was established in ref. \citenum{faribault:2022}, while for RG states of different seniorities this is \emph{the} message of Part I. When the two RG states are the same, $J$ becomes the Jacobian of the equations \eqref{eq:ebv} and is generally non-singular. Its scaled first cofactors are explicitly the elements of the transpose of $J^{-1}$, while its scaled higher cofactors are determinants of its scaled first cofactors. This is a theorem of Jacobi.\cite{vein_book} Inverting $J$ thus provides all the information for the matrix elements of one RG state. On the other hand, when the two RG states are different (and any blocked orbitals are identical), $J$ is necessarily singular: its determinant represents the overlap of two distinct eigenvectors of $\hat{H}_{BCS}$. In this case, the approach of Chen and Scuseria from ref. \citenum{chen:2023} may be employed: the singular value decomposition of $J$ isolates the singularities in its inverse, which when scaled by $\det(J)$ removes them entirely. Higher cofactors are computed in the same manner.\cite{johnson:2024b} When the two RG states have different seniorities or distinct blocked orbitals the matrix $J$ is usually but not always singular. In any case, a single linear algebra operation is required to compute all the matrix elements between two particular RG states. 

It has been shown that a single RG state may be employed as a variational wavefunction, giving a result close to seniority-zero CI for bond-breaking processes.\cite{fecteau:2020} The variational parameters to optimize $\{\varepsilon,g\}$ are those defining $\hat{H}_{BCS}$, providing not one but a complete set of RG states. Naturally, a notion of excitation level between RG states comes from comparing their bitstrings: a single excitation exchanges one 1 with one 0, while double exchanges two 1s with two 0s, etc. A CI built from the variationally obtained RG reference along with its single and double excitations is completely indiscernible from seniority-zero CI.\cite{johnson:2023,johnson:2024a} But this is unnecessarily expensive: single-reference ENPT2, computed from the RG reference $\ket{\Psi_0}$, and other RG states $\ket{\Psi_{\alpha}}$
\begin{align} \label{eq:enpt}
	E^{(2)}_{EN} = \sum_{\alpha >0} \frac{ \vert \braket{\Psi_{\alpha} | \hat{H}_C | \Psi_0} \vert^2 }{E_0 - E_{\alpha}}
\end{align}
is much more affordable, and matches seniority-zero CI to $10^{-5}\;E_h$. In equation \eqref{eq:enpt}, the energies in the denominator are not those of the model $\hat{H}_{BCS}$, but of the target $\hat{H}_C$ (the system of interest). 

The current goal is to add the contributions from RG states of higher seniorities, particularly seniorities two and four. In Part I, a naive classification was made based only on comparing bitstrings. A CI was built, computed by brute force, and compared with seniority 0+2 and seniority 0+2+4 CI (computed with Slater determinants). While the results were not terrible (maximum deviation of $6\;mE_h$ for linear H$_6$), it was clear that something was missing. Thus Part I ended with the resignation that a tedious systematic study of all couplings between RG states of different seniorities, counting the number of near-zero singular values for each possibility, would eventually lead to an answer. Thankfully, it turns out that the naive classification was nearly correct, missing only some important seniority-two RG states. Before summarizing them, it is necessary to understand the nature of the optimal RG reference state. It is similar to the generalized valence bond / perfect pairing (GVB-PP)\cite{hurley:1953,hunt:1972,hay:1972,goddard:1973} wavefunction: it is a collection of subsystems each containing one pair of electrons principally in two orbitals. In GVB-PP, these subsystems are strictly disjoint, while for RG states they interact, very weakly, with one another. Each subsystem contributes 10 to the bitstring, which becomes $(10)^M$, and the $\{\varepsilon\}$ arrange themselves into groups of two, well-separated from another compared with the pairing strength $g$. The explicit connections between GVB-PP and this RG state will be developed in detail in an upcoming contribution.

Now, to identify which seniority-two and seniority-four RG states matter, the individual ENPT2 energy corrections for \emph{each} state were computed for a reasonably-sized system (10 pairs in 20 orbitals). It immediately became evident that only the states listed in Table \ref{tab:Slater-Condon} give any contribution at all. Of the remaining seniority-two RG states, the largest energetic contribution is on the order of 10$^{-6}\;E_h$, while there are no meaningful contributions from the remaining seniority-four RG states.
\begin{table}[h!]
	\centering
	\begin{tabular}{|c|c|c|c|}
		\hline
		\textbf{Label} & \textbf{Rule} & \textbf{Example} & \textbf{Dimension} \\
		\hline
		local single & 10 $\rightarrow$ XX & 10 10 XX 10 & $M$ \\
		\hline
		typical single & 10 $\rightarrow$ X0 and 10 $\rightarrow$ 1X & 10 X0 1X 10 & $M(M-1)$ \\
		\hline
		atypical single & 10 $\rightarrow$ X1 and 10 $\rightarrow$ 0X & 10 X1 0X 10 & $M(M-1)$ \\
		\hline
		s2 double & two 1 $\rightarrow$ X and 0 $\rightarrow$ 1 & X1 X0 10 10 & $\binom{M}{2} (N-M)$ \\
		\hline
		s2 double & 1 $\rightarrow$ 0 and two 0 $\rightarrow$ X & 0X 10 1X 10 & $M \binom{N-M}{2}$ \\
		\hline
		s4 double & two 1 $\rightarrow$ X and two 0 $\rightarrow$ X & 1X XX 10 X0 & $\binom{M}{2} \binom{N-M}{2}$ \\
		\hline
	\end{tabular}
	\caption{RG states of seniorities two and four which couple to the reference bitstring $(10)^M$ for $M$ pairs in $N$ orbitals. Examples are presented as excitations from 10 10 10 10.}
	\label{tab:Slater-Condon}
\end{table}
The three types of single excitation require the notion of subsystems whereas the double excitations do not. The singles are listed separately as they behave differently from one another. First, while local singles are listed, they do not couple to the RG reference, or if so, only very weakly as a variational orbital optimization specifically eliminates these couplings. But what about the other singles? Consider the bitstring 10 10 10 10. An orbital optimization will not decouple it from the typical single 10 X0 1X 10, but from a linear combination of that typical single \emph{and} the atypical single 10 X1 0X 10. As a result, the atypical singles will be important. The name ``atypical'' is chosen as it appears that more is happening: one 1 becomes X, one 0 becomes X \emph{and} one 0 becomes 1, but they are as important as the typical singles. The double excitations are precisely those suggested by the naive classification. Thus, all that was missing from Part I were the seniority-two atypical single excitations.

With the relevant seniority-two and seniority-four excitations identified, it is time to move on to computations.

\section{Numerical Results}
The dissociation of linear equidistant H$_8$ is presented as it is strongly correlated yet small enough to compare with full CI. All computations were performed in the STO-6G basis, with optimal seniority-zero CI orbitals.\cite{johnson:2020} Seniority-based Slater determinant CI and full CI results were computed with PyCI.\cite{richer:2024}  

\begin{figure} 
	\begin{subfigure}{\textwidth}
		\includegraphics[width=0.485\textwidth]{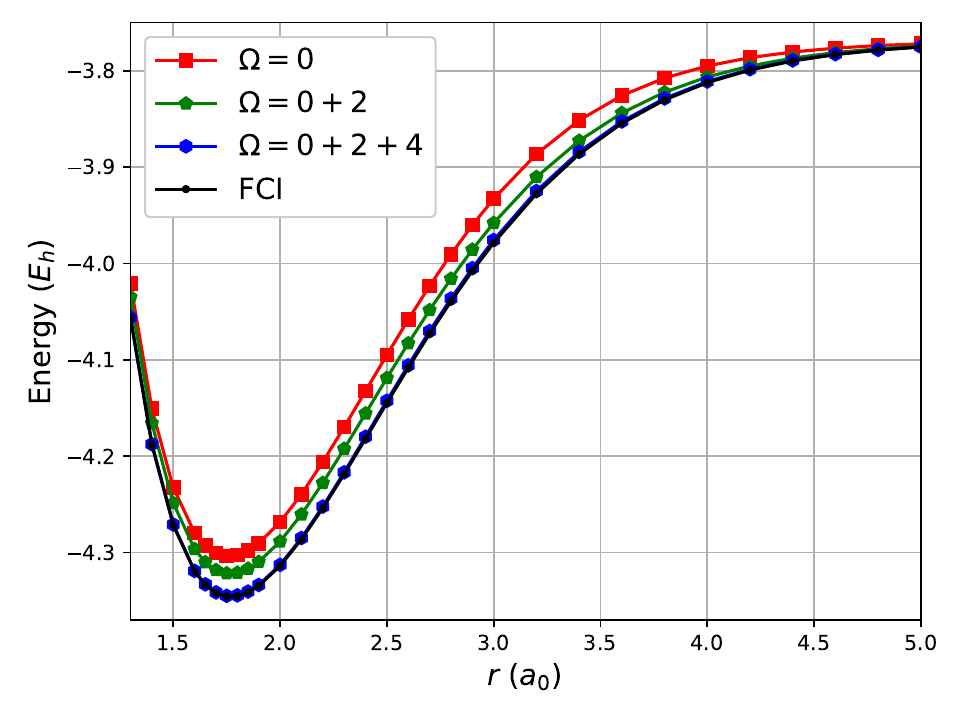}
	\end{subfigure}
	\caption[]{Seniority ($\Omega$)-based CI curves for the symmetric bond dissociation of linear H$_8$. Results are computed in the optimal orbitals for seniority-zero Slater determinant CI obtained in ref. \citenum{johnson:2020} in the STO-6G basis.}
	\label{fig:s024_CI_h8}
\end{figure}

\begin{figure} 
	\begin{subfigure}{\textwidth}
		\includegraphics[width=0.485\textwidth]{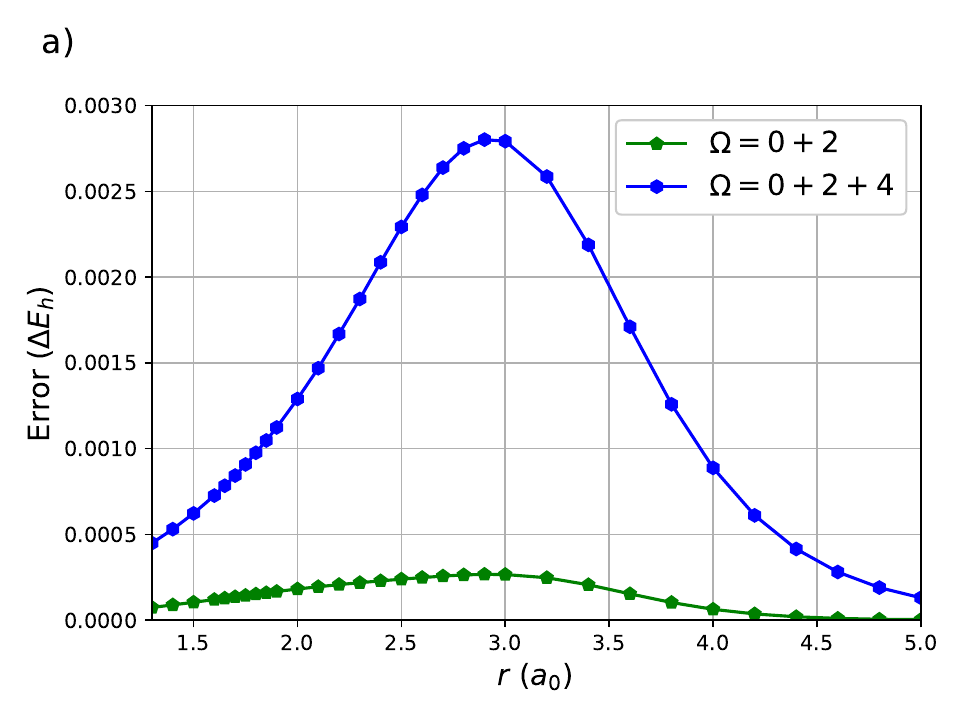}
		\hfill
		\includegraphics[width=0.485\textwidth]{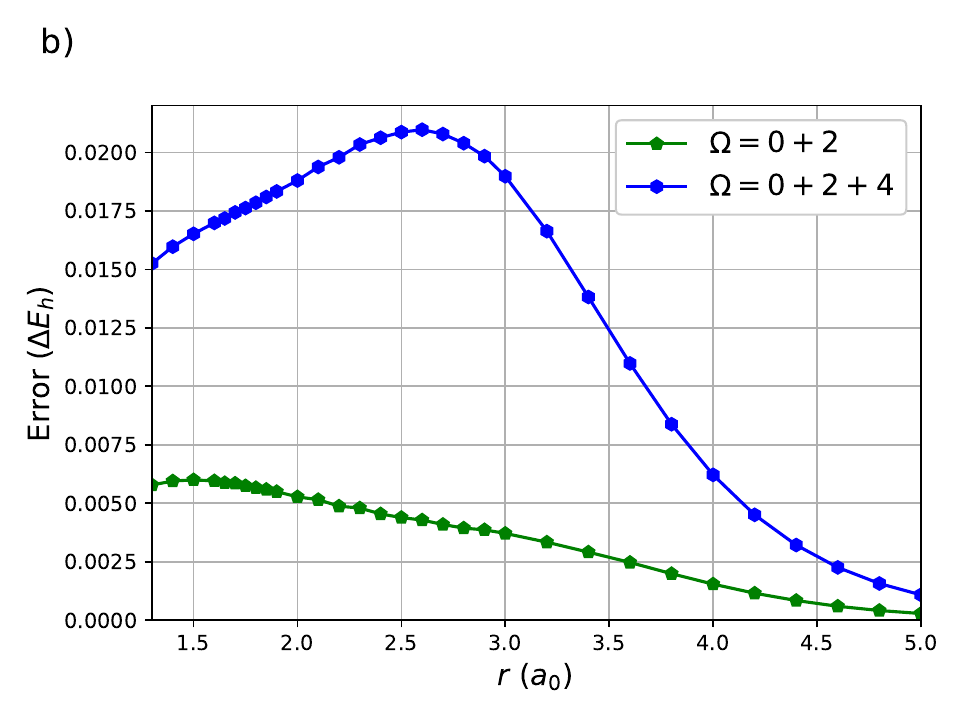}
	\end{subfigure}
	\caption[]{(a) Errors of seniority ($\Omega$)-based RG-CISD curves for the symmetric bond dissociation of linear H$_8$ with respect to seniority-based Slater determinant CI. (b) Errors of seniority ($\Omega$)-based RG-ENPT2 curves for the symmetric bond dissociation of linear H$_8$ with respect to seniority-based Slater determinant CI.  Results are computed in the optimal orbitals for seniority-zero Slater determinant CI obtained in ref. \citenum{johnson:2020} in the STO-6G basis.}
	\label{fig:s024_RG_h8}
\end{figure}

First, figure \ref{fig:s024_CI_h8} 
demonstrates the rapid convergence of seniority-based Slater determinant CI for linear H$_8$. These results will be approximated with two different methods, both using one RG reference and its single and double excitations. The easiest and cheapest approximation is perturbative, specifically single-reference ENPT2 \eqref{eq:enpt}, which only computes the couplings of each RG excited state with the given reference. At the other extreme, explicitly constructing and diagonalizing the Hamiltonian in the collection of states is the most expensive approach. While in the seniority-zero sector these approaches yielded near identical results,\cite{johnson:2024b} one can see in figure \ref{fig:s024_RG_h8} 
that this is not the case here. From figure \ref{fig:s024_RG_h8} (a), RG-CISD is excellent in seniority 0+2, and acceptable in seniority 0+2+4, where it has a maximum deviation on the order of 3 $mE_h$. Figure \ref{fig:s024_RG_h8} (b) shows that RG-ENPT2 is acceptable in seniority 0+2, but unacceptable in seniority 0+2+4 where it deviates by more than 20 $mE_h$. For both methods, the maximum disagreement occurs in the recoupling region between equilibrium (weak correlation dominant) and dissociation (strong correlation dominant). Both methods dissociate the H atoms correctly. 

There are two points to underline. First, that partitioning the Hilbert space in terms of one RG state and its weak excitations is effective: in each seniority sector, RG-CISD is a good approximation. Second, the choice of cheap and expensive methods using the same set of states was intended to draw lines around the amount of effort required. The RG-CISD matrix has a size of $\mathcal{O}(M^4)$, so that its diagonalization costs $\mathcal{O}(M^{12})$. While this is a polynomial, clearly this is not desirable. On the other hand, RG-ENPT2 is cheaper, but the results are not good enough. Its cost is the number of states times one linear algebra operation (either matrix inversion of singular value decomposition), which here means $\mathcal{O}(N^7)$. This is also rather high, and should be improved. The tasks going forward are thus clear: to develop an approach intermediate between RG-ENPT2 and RG-CISD, and to develop practical approximations to RG states to bring down their computational cost. 

\section{Conclusion}
Slater determinant CI partitioned by the number of unpaired electrons (seniority) is known to describe strongly correlated systems, though each seniority sector has cost that grows faster than exponentially. In this contribution, it is demonstrated that single-reference approaches may be employed for the same purpose with Richardson-Gaudin states. The relevant states are identified, and using the expressions developed in ref. \citenum{johnson:2025b}, computations demonstrate that a CI in terms of these states converges quickly in each seniority sector. This treatment is expensive, but polynomial, opening the door to development of feasible approximations.

\section{Acknowledgments}
P. A. J. was supported by NSERC and the Digital Research Alliance of Canada.

\section{Data Availability}
The data that support the findings of this study are available from the corresponding author upon reasonable request.

\section{Conflict of Interest}
The author has no conflicts of interest to disclose.

\bibliography{non_zero_sen_2}
\bibliographystyle{unsrt}

\end{document}